# Super-resolution imaging and optomechanical manipulation using optical nanojet for non-destructive single- cell research

Alina Karabchevsky[1,*], Tal Elbaz[1], Aviad Katiyi[1], Ofer Prager[2], Alon Friedman[2]

[1] School of Electrical and Computer Engineering, Electro-Optics and Photonics Engineering Department, Ben-Gurion University of the Negev, Beer-Sheva, Israel.

[2] Faculty of Health Science, Ben-Gurion University of the Negev, Beer-Sheva, Israel.

* Corresponding author. E-mail: alinak@bgu.ac.il

Advanced photonic tools may enable researchers and clinicians to visualize, track, control and manipulate biological processes at the single-cell level in space and time. Biological systems are complex and highly organized on both spatial and temporal levels. If we are to study, perturb, engineer or heal biological entities, we must be able to visualize key players in such systems and to track, control and manipulate them precisely and selectively. To achieve this goal, the engineering of non-destructive tools will allow us to interrogate and manipulate the function of proteins, pathways and cells for physicians, enabling the design of 'smart materials' that can direct and respond to biological processes on-demand. Among the potentially exploitable non-destructive tools, light-based actuation is particularly desirable, as it enables high spatial and temporal resolution, dosage control, minimal disturbance to biological systems and deep tissue penetration. Here, we overview existing approaches toward the engineering of light-activated tools for the interrogation and manipulation of single-cell processes, and list the types of studies and types of functions that can be controlled by light. Timely applications, such as studies of inflammation and of crossing brain barrier systems - via super-resolution imaging and optomechanical manipulation - are two representative examples of emerging applications so far never addressed.

# 1 Introduction

Existing methodologies in the engineering of light-activated tools limit the types of studies and functions that can be controlled by light for the interrogation and manipulation of single-cell processes.

Understanding biological systems such as the *origin*, *evolution*, *development* and *prognosis* [1] of diseases is dictated by processes within a **single cell**, therefore research at the single-cell level is an important field in molecular biology and medicine. The first and most used approach employed is to closely view a sample through the magnification of a lens, using visible light (Figure 1a). The wave nature of light limits the 3D resolution in classic optical microscopy. In 1873, Ernst Abbe described the resolution limit of an optical microscope [2]. Using wave optics, Abbe showed that the limit in the ability of a conventional optical microscope to resolve an object is $dx = k\lambda/NA$, where $d$ is the distance between two point sources, and NA is the numerical aperture, $NA = n \sin \theta$ where $\theta$ is the half-angle of the light cone that can enter the lens (objective) from a point source, and $n$ is the refractive index of the object-space. Depending on the definition, the coefficient $k$ is in the range of 0.473-0.61 and the resolution is in the range of 200-500 nm. However, optical microscopes are unable to reach a theoretical optical resolution due to aberrations and imperfections in classical optics, and also suffer from low contrast images captured from cells [3, 4, 5].

The second approach, developed in the 20th century and widely used for the studies of dynamics and localization of intra- and extracellular components in cells, utilizes organic fluorescent dyes [2, 6, 7, 8].

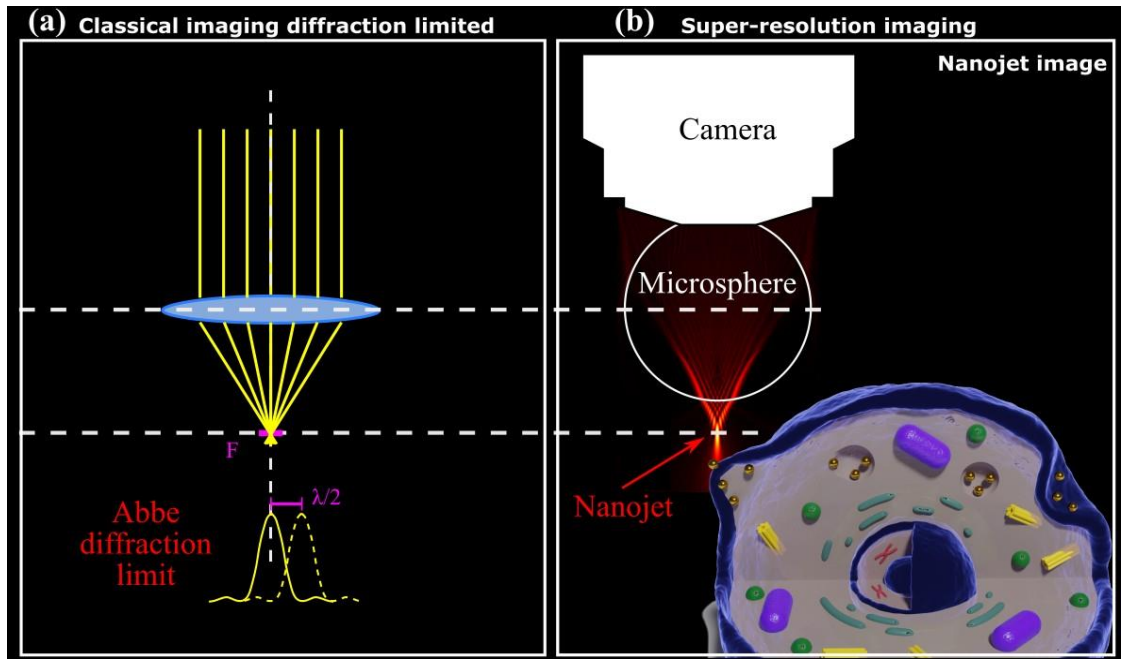

Figure 1: Concept of classical imaging as compared to super-resolution imaging. (a) Classical imaging microscopy with a conventional lens where the system is limited by Abbe diffraction limit as illustrated below. (b) Illustration of super- resolution imaging by microsphere with nanojet.

However, utilizing naturally light-responsive fluorophores to engineer light-mediated control over a diversity of proteins or functions can be difficult or impossible, highlighting the most significant drawback of this approach, which is its lack of generality and its inability of sustaining a long-lasting fluorescence due to fast photobleaching. Because of the light stability (photostability), fluorescent semiconductor quantum dots (QDs) have been proposed as an alternative to organic fluorescent optical probes [9]; however, they cannot be used in live cells due to their cytotoxicity and strong blinking.

The third approach was developed in the early 21st century and provides super-resolution imaging in live cells [10]. The most common method for super resolution techniques is Stimulated-emission-depletion fluorescence (STED) microscopy [11, 12]  This method is based on selectively deactivating the fluorescence with a doughnut-shaped beam and confining it as a smaller point. It can achieve 50-60 nm lateral resolution [13]. Another method is single-molecule localization microscopy (SMLM) [14, 15] which is based on localizing an individual fluorescent molecule. SMLM can achieve high spatial resolution – typically 20-50 nm. This method is used in photoactivated localization microscopy (PALM) [16] and stochastic optical reconstruction microscopy (STORM) [17]. In addition, super-resolution can be achieved in a label-free manner. One method for high resolution is coherent anti-Stokes Raman scattering (CARS) microscopy [18], which is a nonlinear optical version of Raman scattering. The elastic scattering which appears in Raman scattering is canceled by illumination with a pump and Stokes beams, forcing only inelastic scattering (anti-Stokes) to occur. This method provides high resolution and fast imaging. Another method is based on phase difference and is called quantitative phase imaging (QPI) [19, 20]. QPI is a very powerful tool for free-label imaging because it combines microscopy, holography and light scattering techniques. It can be used for imaging completely transparent structures. Substantial drawbacks of this approach are 1) that the smallest feature it can detect is in the level of noise that stems from the fluctuation in the data, and 2) incompatibility with conventional camera technologies such as CMOS cameras. In addition, this method is inherently irreversibly destructive and, once applied, the protein or molecule cannot revert to its original state. Thus, this technology is limited to a one-time-only usage that imposes a substantial restriction on the types of studies and applications that can be advanced with this methodology.

The fourth approach was developed for 3D optical manipulation of microparticles, cells and biomolecules in a non-contact and non-invasive manner, by the use of optical tweezers [21, 22]. Despite being a widely used optical manipulation technique, it is limited in precise manipulation in biological applications because of its bulky lens system and limited penetration depth. Since the diameter of biomolecules is in the range of 1-10 nm (small molecules such as IL-1$\beta$ are in the range of 4-7 nm), the optical tweezer cannot directly manipulate them due to the diffraction limit of light. Emerging near-field methods of plasmonic tweezers and photonic crystal resonators can overcome these limitations; however their heating effect may damage biological specimens [23]. In this review, we summarize the development of advanced photonic tools for biological research at a scale of single-cell investigation. The paper is structured as follows: Section 2 is focused on the super-resolution configurations that allow visualizing tiny changes in a cell. Section 3 describes the fundamental formation of optical forces. Section 4 elaborates on the utilization of gold nanoparticles for single cell investigation. Section 5 is focused on the computational methods to design the optical system and analyze its output. Section 6 shows two representative examples of emerging applications.

## 2 Super-resolution imaging

To trace and view the biological processes such as endocytosis and exocytosis of nanoparticles encapsulated by small molecules such as IL-1$\beta$ ligands by macrophage cells or others, one can utilize the photonic nanojet effect (Figure 1b). Photonic nanojet is one of the approaches developed to break the postulate dictated by Abbe in classical optics (Figure 1a). One such approach is the photonic nanojet effect [24, 25, 26].

A photonic nanojet is a narrow light beam situated near the shadow-side surface of an illuminated dielectric microsphere, whose diameter is comparable to one or a few wavelengths of the light source [26]. The nanojet exhibits: 1) high intensity (up to x100 the incident power density); 2) subdiffraction beamwidth; 3) several wavelengths long reach. In 2009, J. Y. Lee et al. observed that self-assembled nano- or micro- lenses can resolve features beyond the diffraction limit using an ordinary white-light microscope [27]. In 2011, Z. B. Wang et al. reported that dielectric microspheres illuminated with white light and combined with a standard microscope are capable of reaching resolution features as small as 50 nm resolution integrated with a standard microscope [28]. Such a remote-mode microsphere nano-imaging system can allow a working distance larger than the wavelength of the light source. While these studies illustrate the feasibility of breaking the fundamental Abbe limit to achieving super-resolution imaging, the technologies used have a significant drawback: The described auxiliary structures such as self-assembled nano- or micro-lenses or dielectric microspheres operate in air, which limits the general applicability of these technologies to biological applications. Figure 2 shows the super-resolution imaging setups and the imaging outcomes for visualization at a single cell level. The common method utilizes a microsphere for super-resolution as shown in Figure 2A. This method uses a microsphere as magnifier and can achieve a super-resolution of 680 nm [29]. Another method for biological application is by using bio-cell magnifiers as shown in Figure 2B. These biomagnifiers enable a resolution of up to 100 nm and prevent mechanical and photothermal damage to biospecimens [30].

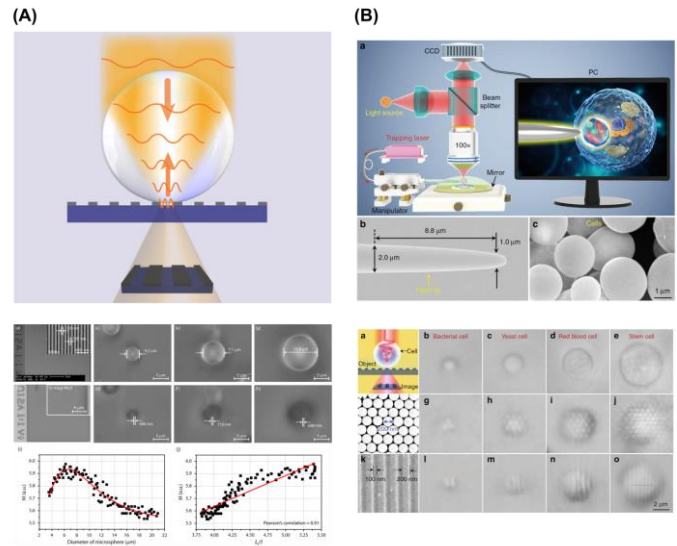

Figure 2: Super resolution imaging: (A) Magnification by a dielectric microsphere (reproduced from [29]). (B) Super resolution by biomagnifiers. Top line shows schematic of the experimental setup uses for attaching the biospecimens to the probe fiber. Bottom line shows different resolutions of different objects with a variety of biomagnifiers (reproduced from [30]).

Magnifiers can be engineered from biocompatible $BaTiO_3$, while the auxiliary structure is of a cylindrical shape. Figure 3 shows photonic jet formation via 12 $\mu$m $BaTiO_3$ microcylinder with refractive index $n_{cylinder}$ = 2.4-2.5 in water (refractive index $n_{water}$ = 1.33) for two for 520 nm and 637 nm respectively. The refractive index contrast between water and $BaTiO_3$ is high enough and is therefore suitable for biological environments [31].

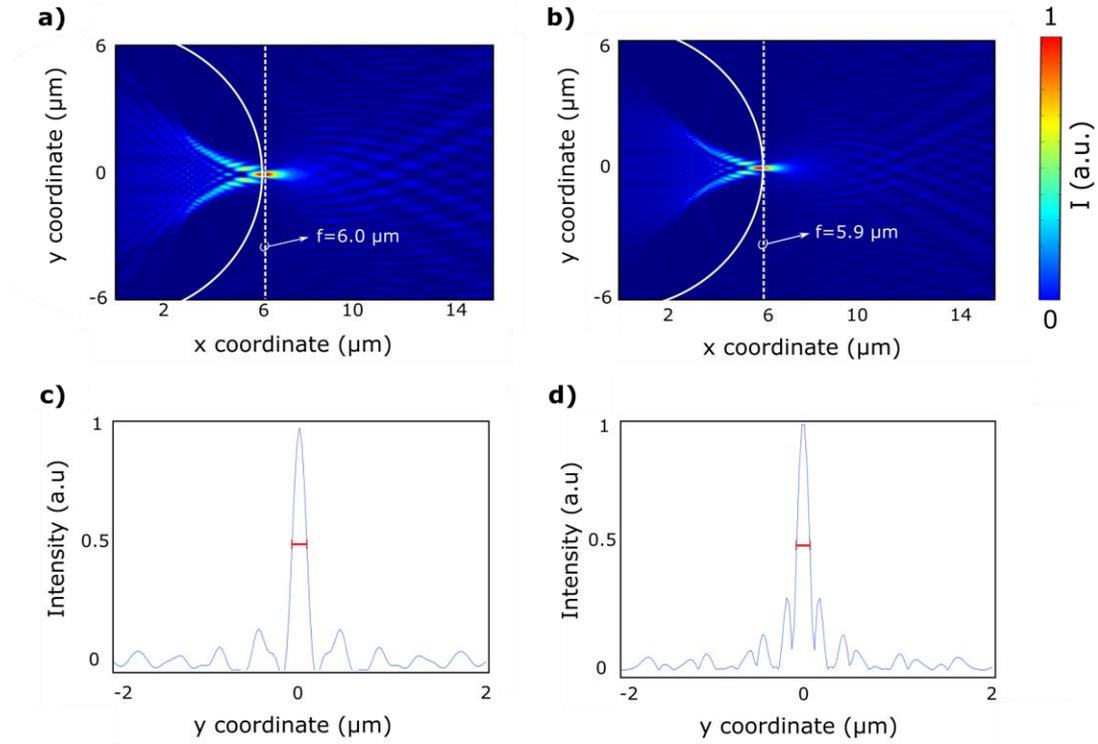

Figure 3: Calculated photonic jet out of BaTiO$_3$ microcylinder with diameter of 12 micron in water medium illuminated at wavelength of (a) 520 nm and (b) 637 nm. Normalized distribution of intensity for wavelength of (c) 520 nm and (d) 637 nm.

Table 1 shows calculated results of key parameters of a photonic Jet formation via an illuminated 12 µm BaTiO3 microcylinder auxiliary structure in water. The FWHM values demonstrate super-resolution values beyond the diffraction limit, which is half of the wavelength. The focal point of maximum intensity is measured from the microcylinder, is formed at the near-field, and enables the effect of narrowing of the photonic jet that exists only near the particle surface [32].

Table 1: Key parameters of a photonic jet generated via 12 $\mu$m BaTiO$_3$ microcylinder in water for two different wave- lengths in visible:

| Wavelength [nm] | Focal Length [$\mu$m] | Max Intensity [W/m$^2$] | FWHM [nm] | Longitudinal length [$\mu$m] |
|---|---|---|---|---|
| 520 | 5.9 | $68.2 \times 10^{-3}$ | 0.30$\lambda$ | 1.3 |
| 637 | 6.0 | $51.4 \times 10^{-3}$ | 0.22$\lambda$ | 1.6 |

# 3 Optical forces

## 3.1 Fundamentals of Optical Forces

Shortly after the laser was invented, Ashkin proposed and demonstrated [33] that the incident laser beam can be implemented to trap and manipulate dielectric particles and other cells.

The optical force on mesoscale particles of micro- or nano-scale is a consequence of the conservation law of electromagnetic wave momentum. At these sizes, the particle does not correspond to the gravitational force, but to the radiation pressure and the Lorentz force. Theoretical modeling of optical forces was discussed by Debye back in 1909 [34, 35], showing that the force applied on a sphere can be described in terms of extinction and scattering cross-sections as a function of microsphere dimensions with respect to the beam wavelength. Ashkin, back in 1970, set the foundations of optical trapping of micron-sized particles resulting from the radiation pressure from intense, coherent laser [36, 37]. For this discovery he was awarded the Nobel prize in 2018 [38]. The radiation pressure, expressed by the momentum carried by a single photon $p$, has two different approaches of calculation, known as the Abraham-Minkowski con- troversy [34]:

$$p_{\text{Min}} = \hbar k = \frac{n\hbar\omega}{c} \qquad (1)$$

or

$$p_{\text{Abr}} = mv = \frac{\hbar\omega}{nc} \qquad (2)$$

where $\hbar$ is the reduced Planck's constant, $\omega$ is the angular frequency of the light, c is the speed of light in a vacuum and $n$ is the refractive index of the medium. Barnett suggested that $p_{\text{Min}}$ describes the canonical momentum, while $p_{\text{Abr}}$ describes the kinetic momentum. When considering Minkowski's force theory and Abraham's force theory as time-averaged forces, the stress tensor coincides for both. In the case of a Rayleigh particle, when the object is much smaller than the wavelength it behaves like an electric dipole [36, 39]. The Lorenz-Mie theory, among others, is an analytical method designed for plane-wave scattering by a spherical particle. Barton and co-workers added corrections to the fundamental Gaussian beam, enabling the derivation of the incident and scattered fields from a sphere [40]. In this way the force can be calculated by means of integration of the Maxwell stress tensor **T** which is the time-averaged optical force acting on a particle over a closed surface $\partial V$ surrounding the particle [30, 41]:

$$\langle \mathbf{F} \rangle = \int_{\partial V} \langle \mathbf{T}(\mathbf{r}, t) \rangle \cdot \mathbf{n}(\mathbf{r}) \, \mathrm{d}A. \tag{3}$$

where **T** is the time averaged Maxwell stress tensor and **n** is the normal.

We generally refer to the particle being moved around by this force as the 'probe' or 'target', bearing an arbitrary rigid shape and size. However, for numerical simulations, this method of obtaining optical forces is computationally intensive. For instance, to obtain the full force field over a two-dimensional space we would need to place the probe particle in one specific point in space, evaluate the electromagnetic field in this configuration, and then integrate over the surface around the probe to obtain the forces, repeating the process over a 2D grid of points. Additionally, the simulation in this method needs to be performed using the full 3D geometry.

Assuming that our probe is within the Rayleigh regime (i.e., the incident wavelength is much larger than probe dimensions), we can assume that the optical force acting on the target can be described using the Taylor expansion. Physically, this would imply that the probe is approximated to only have electric and magnetic dipole moments. The optical force equation becomes [41, 42]

$$\langle \mathbf{F} \rangle = \frac{1}{2}\mathrm{Re}[(\nabla \mathbf{E}_i^*) \cdot \mathbf{p}] + \frac{1}{2}\mathrm{Re}[(\nabla \mathbf{B}_i^*) \cdot \mathbf{m}] - \frac{k^4}{12\pi\varepsilon_0 c}\mathrm{Re}[\mathbf{p} \times m^*] + \cdots \tag{4}$$

We deal with only the first term of Eq. (4), which corresponds to the force if the probe were assumed to be only an electric dipole. The second term (the force if we were also to consider a magnetic dipole term in the Taylor series expansion), the third term (the force due to the interaction of the two dipoles), as well as all higher-order terms, are negligible and can be ignored. This approximation also greatly simplifies numerical simulations, as we need to obtain only the electromagnetic fields on one plane and the domain can also be simplified into 2D geometry (whenever applicable). The first term of Eq. (4) can be written in the following form:

$$\langle \mathbf{F} \rangle = \frac{\alpha'}{4}\nabla E_0^2 + \frac{\alpha''}{2}E_0^2 \nabla \phi \tag{5}$$

where $\alpha$ is the probe's complex polarizability. For a spherical particle, this is given by [43]

$$\alpha = 4\pi r^3 \varepsilon_0 \frac{\varepsilon_p - \varepsilon_a}{\varepsilon_p + 2\varepsilon_a} \qquad (6)$$

We note that the value of the polarizability is dependent on the difference of the dielectric permittivities of both the probe particle $\varepsilon_p$ and the surrounding material $\varepsilon_a$. As described in Figure 4, the first term in Eq. (5) is the *gradient force*, which arises from field inhomogeneities and is proportional to the dispersive interaction of the induced atomic dipole with the intensity gradient of the light field [44], whereas the second term is the *scattering force*, which can be regarded as a consequence of momentum transfer from the radiation field to the particle, proportional to the dissipative (imaginary) part of the complex polarizability [41]. Note that the phase $\varphi$ can be written in terms of the **k** vector such that $\nabla \varphi = \mathbf{k}$.

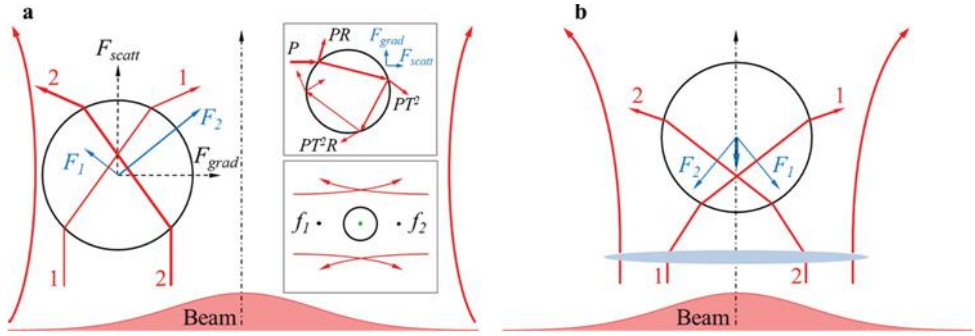

Figure 4: Optical forces acting on a dielectric microsphere illuminated by a Gaussian beam. (a) Total forces resulted from scattering force $\mathbf{F}_{scatt}$ and gradient force $\mathbf{F}_{grad}$ and (b) net forces resulted from focused rays (reproduced from [36]).

For certain biological applications, such as label-free manipulation of cells, the polarization $\alpha$ is very low due to the low permittivity contrast of the probe and the environment; hence a high-power and highly- focused incident field is required for this application. This then raises the problem of heating. In addition to the potential for altering or damaging the biological samples, heating can also influence the motion of the targets due to thermophoretic effects [45]. One method of solving the contrast problem is to add gold nanoparticles in the cells to serve as 'handles' which can influence the movement of the entire system. As a solution, one can use the photonic hook to move cells in their liquid background [46], as detailed in the following subsection.

## 3.2 Optomechanical manipulation: moving objects in inter- and intra- cellular media

To move objects in inter- and intra-cellular media by the use of light, one needs to create an optical force (see preceding section). Based on the properties of the electromagnetic field, optical forces can be classified into two main categories, namely, scattering and gradient forces. The capability of light to exert forces on particles, a phenomenon known as *radiation pressure*, is well-known and extensively documented in the literature. Historically, demonstrations of radiation pressure showed that this force is very small, but with the invention of the highly directed and focused laser beams, control over the position of small objects can be realized. The act of moving objects with the help of electromagnetic forces is termed 'optical tweezers'. Optical tweezers have found applications in biology and nanotechnology; however, they come with several limitations. Highly focused laser beams require expensive, complicated, and bulky lens set-ups. Multiple traps can be formed through different methods that require increased laser power input, etc.

Producing mechanical action on particles through electromagnetic radiation was first hypothesized from the observation that the direction of a comet tail points away from the sun. Kepler suggested that this phenomenon is a result of radiation pressure, and Maxwell [47] later showed that momentum transfer from the EM field to an object result in this radiation pressure. Specifically, he showed that the force on an object absorbing $P$ watts of light is given by $F = P/c$, whereas for a perfectly reflecting object, this force is $F = 2P/c$.

Lebedev [48] and Nichols [49] independently showed this effect experimentally for macroscopic objects in the years 1901 and 1903, respectively. Additionally, Lebedev also studied optical forces on gases in 1910 [50]. By manipulating the field, Ashkin [51, 52] showed experimentally that it is possible to trap particles at the field focal point, showing that control of the field can be used to re-position objects, and even to move particles opposite to the direction of the incident field [53, 54, 55].

## 3.3 Photonic Jet and Photonic Hook formation

The conventional approach for generating an optical force utilises high magnification objectives for focusing the laser beam as shown in Figure 5A, for instance for levitating particles [56], as shown in Figure 5B. While the traditional microscope-based trap has found applications in a wide variety of disciplines, such as biophysical research, the light-based approach is diffraction limited. Thus, achieving manipulation on the nanoscale requires auxiliary structures that generate tightly confined electric fields.

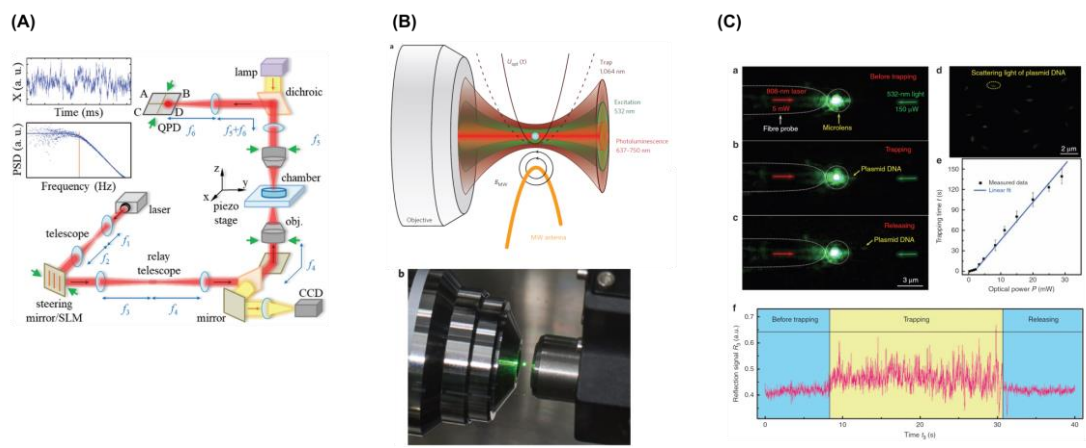

Figure 5: Examples of optical forces exerted on a cell or particle: (A) Conventional optical tweezers system (reproduced from [36]). (B) **a** A hybrid nano-optomechanical system and **b** a levitated nanodiamond (reproduced from [56]). (C) Opti- cal tweezers for biosamples by nanojet with a probe fiber (reproduced from [68]).

The *photonic nanojets* (PNJs) are non-evanescent and low divergence localized spots with high intensity [57] that are observed on the shadow side of the surface of the auxiliary structure [58] (spherical or square [59], cylindrical [60] or other shapes [61, 62]) when illuminated by a plane wave [63] or pulsed illumination [61]. The PNJ is generated at the near-field due to a wavefront deformation resulting from interactions of the low absorbing microelement boundary and the illuminating field. According to Mie theory, the optical fields inside and outside the dielectric microelement are subject to the light wave, focusing it and acting as a refractive microlens due to the curvature of the particle surface. It has negligible aberration and high NA when the refractive index contrast between the microsphere and its surrounding medium is about 1.5 [30,

58, 64, 65]. The PNJs are characterized by a full width at half-maximum (FWHM), capable of going beyond the diffraction limit [66]. Due to their unique properties, they can be utilized for super-resolution imaging, photolithography at nanoscale, and biological applications, using nanoparticles [65]. The key parameters of photonic nanojets are focal distance, longitudinal size, maximum intensity and a lateral FWHM (see Table 1) which depends strongly on the refractive index contrast and auxiliary structure dimensions [58, 65, 67].

The actual generation of PNJs is usually realized with a plane wave incident on spherical or cylindrical nanoparticles [69, 70, 60], but PNJs also generated from non-spherical particle cross-sections such as in refs. [62, 46, 71, 61, 72, 66]. In addition, photonic jets have been previously used for optical manipulation applications, the first of which was done by Cui, et al., [73], where they obtained the forces on a metallic nanoparticle immersed in the PNJ field. This way they can be used for manipulation and detection of biological targets [74].

Photonic nanojets, discussed above, can also be generated using other symmetric structures. In the con- text of optical manipulation at a single cell level, the continuous wave (CW) field-based illumination causes heating to the system, which can be potentially destructive. To overcome the destructive operation of CW, an alternative approach of pulsed laser-generated input field [61, 75] was proposed. It was shown [61] that this approach also increases the optical force performance, which has been demonstrated experimentally with femtosecond lasers while trapping latex nanoparticles [76]. Because of its conservative character, the optical force can be derived from potential energy, the minima of which can be used for light-activated motion. Optical forces are usually generated using spherical structures subjected to the illumination of a plane CW. When optical forces in the form of photonic nanojets are applied on a metallic nanoparticle, the forces acting on this nanoparticle are the result of the momentum exchange [77].

Worth noting is that these previous works are primarily focused on trapping the particles along some axis of symmetry. Curved photonic jets, named *photonic hooks*, can be used for creating forces that move an object in inter- or intra-cellular media in a curved trajectory. These photonic hooks are made possible using asymmetric particles [80, 81]. By breaking the symmetry of an auxiliary object, the generated structured light beams become curved. This effect is known as a curved photonic jet or photonic hook (PH) [79, 71, 82, 83] and has unique properties of a tilt angle depending on deformation [60]. Specifically, PH experiences both lateral size and a radius of curvature as a fraction of the incident wavelength [83]. The distinctive features of the PH are that the transverse size and curvature radius of the beam are a fraction of the incident wavelength, and that the side lobes differ from the shape of the main beam and so do not bend. The generation of well-studied curved Airy beams usually requires expensive and complicated optical elements with a cubic phase, which makes optical elements incompatible with optical systems. The geometries of photonic hook systems are shown in Figure 6. The simple method for breaking the symmetry that will create a photonic hook is by breaking the symmetry of illumination, as shown in Figure 6A. This can be done via blocking part of the input light of an auxiliary structure, as in ref. [60]. By varying the thickness of the blocked beam, for instance with a mask, one can tune the angle of the hook. Another method is to break the symmetry of the focusing object itself. By locally changing the refractive index of the cylinder one can break the system symmetry [78], as shown in Figure 6B. Another method is by fabricating a non-symmetric structure [79] as shown in Figure 6C.

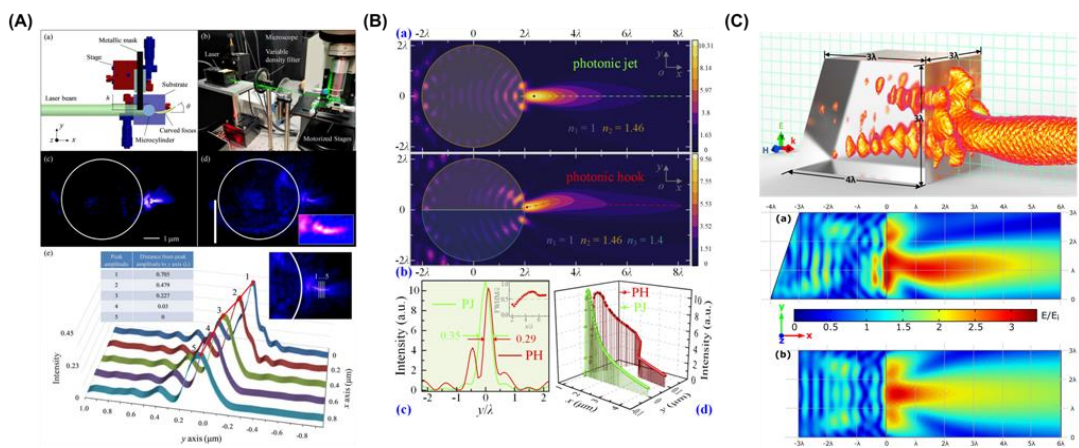

Figure 6: Configurations for photonic hook: (A) A dielectric microcylinder with cover by a metallic mask (reproduced from [60]). (B) Asymmetric microcylinder made by two different materials (reproduced from [78]). (C) A symmetric prism (reproduced from [79]).

Configurations for PHs have been optimized [84], investigated for optomechanically moving nanoparticles around obstacles [79], and recently experimentally demonstrated [60]. Previously, PH-based optical manipulation under CW illumination was explored, but the generated force has a small magnitude for realistic applications [79]. As mentioned above, one way to mitigate the destructive effect of CW illumination is by utilizing a pulsed input field [61].

Despite rapidly progressing advances in engineering tools activated by light and operating optomechanically in biological media such as optical tweezers, only a small force can be generated, due to fundamental limitations of current methods involving CW light. The approach for engineering light-activated tools is by the generation of optical forces - a class of forces known to experience non-destructive light- activated motion - among them optical tweezers, Airy beam, photonic nanojet and photonic hook [84]. The first property of interest to us is that the photonic nanojet (PNJ) and photonic hook have enabled the engineering of motion tools for sorting, PNJ surgery and others, overcoming the obstacles of light scattering and absorption [85]. The second property of interest to us is that optical forces provide a fundamental property of light to develop a practical platform to study processes within the cell, overcoming the need for fluorescent staining [86, 87, 88]. This property allows natural label-free characterization of biological processes. Recent studies illustrate the feasibility of utilizing optical forces for instance: in vesicles (size tens to thousands of nanometres) [89, 90], in manipulating and dragging a selection of liquid domains in lipid bilayers [91], in trapping viruses (thousands of nanometres) [92], and in trapping [21, 22, 93], sorting [94] and isolation [95] of bacteria and living cells. However, the technologies used have one significant drawback, which is the small optical force able to attract or repel an object, thereby rendering these not feasible for actual applications in cell biology research. The scattering force is directly associated with the wavevector of light and is interpreted as the momentum interchange between light and objects when the propagation path is altered owing to discontinuities in the refraction index. The gradient force essentially refers to the gradient of the field energy intensity, which plays an important role in forming traps by overcoming the scattering force.

## 4 Plasmonic gold nanoparticles for local treatment

Plasmonic nanoparticles are widely used in biological research for drug delivery, temperature assisted treatment and other procedures.

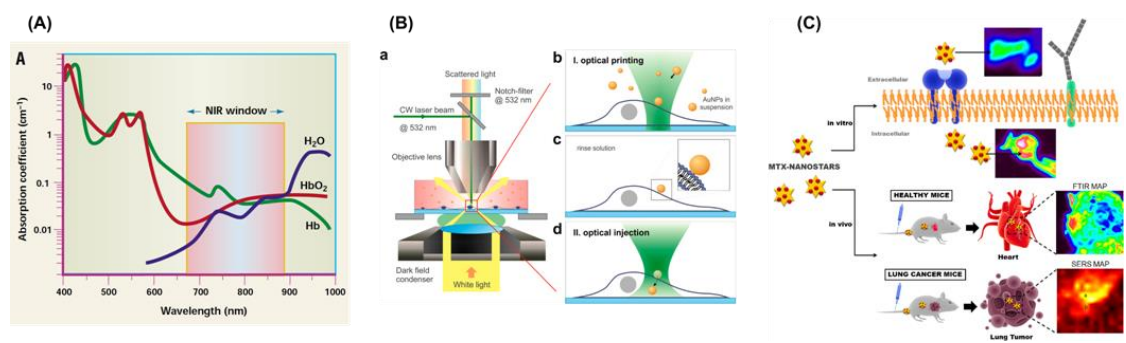

Figure 7: (A) Human tissue transmission window in the near-infrared (reproduced from [96]). (B) Illustration of the optical force system for injecting gold nanoparticles into a living cell (reproduced from [96]). (C) Gold nanostars for efficient drug delivery and monitoring of cancer cells (reproduced from [97]).

Due to their highly polarizable conduction of electrons, gold nanoparticles (NPs) can be used for interaction with the electromagnetic field. In addition, plasmons in gold NPs are excited at 600-900 nm, allowing deeper penetration, since the cells have low absorption at visible wavelengths [98] as shown in Figure 7A. In 1996, the first demonstration used gold nanoparticles to show that insertion of a protein into rat kidney cells [99]. Subsequently, gold nanoparticles were injected into a living cell by using a focused laser beam [96] as shown in Figure 7B. Due to the shift in the GNP plasmon absorption, the change in the placement of the NPs can be verified. The capability of gold NPs to be directed to a specific location can be utilized for NP targeted drug delivery. Since it is relatively straightforward to attach to the gold via a thiolated tail and other methods, it is possible to efficiently control the drug concentration with gold NPs by laser radiation [97], as shown in Figure 7C. Plasmonic gold NP can also be used for imaging. By changing the particle shape, the imaging quality can be improved. For example, by utilizing the Purcell effect, nano-antenna and SPR effect, gold nanorods can be used for plasmon-enhanced fluorescence (PEF), for weak fluorescence emission, and even for single-molecule detection, with an enhancement factor of up to 1100 [100]. Table 2 shows widely used plasmonic nanoparticles for optomechanical manipulation such as nanospheres, nanostars and nanorods. Spherical geometry of nanoparticles is used for drug delivery based on membrane perforation, with typical localized surface plasmon excitation at visible wavelengths [71]. However, spherical particles made of

Mxene, such as in ref. [61], are capable of exciting surface plasmon at longer wavelengths in near-infrared. Nanostars are useful for imaging applications due to multiple hot spots [97], while nanorods are capable of exciting LSPR along the longitudinal and transverse directions, as in refs. [100, 101, 102, 103].

Table 2: Widely used nanoparticles for optomechanical manipulation.

| Shape | Size, nm | Application | Wavelength, nm | Ref. |
| --- | --- | --- | --- | --- |
| Spheres | 5-150 | Cellular medium, drug delivery, optical injection, membrane perforation photonic hook, photonic jet, LSPR | 520-650 ,1550 | [71, 61] |
| Nanostars | 30-40 | Drug delivery and enhanced imaging | 786 | [97] |
| Nanorods | 58x25 | Single molecule enhanced fluorescence | 650 | [100] |

## 5 Computational approaches

To design an optical system for super-resolution imaging or optomechanical manipulation, but also to be able to analyze the obtained experimental results, one may use numerical approaches alongside analytical methods [104, 40]. If the structures are divided into small sub-domains, numerical methods can efficiently handle complex geometries [104]. Popular and efficient computational approaches used for nano-optics and near-field optics include the finite-element method (FEM), the finite-difference time- domain (FDTD) technique, and the finite integral technique (FIT) [61, 105, 104]. The finite-difference time-domain (FDTD), a commonly used numerical modeling tool, is based on an algorithm for discretization of the selected computational domain by proposing a spatial rectangular grid, according to Yee's algorithm back in 1966 [106, 107]. Both electric field $E$- and magnetic field $H$- components are used in a three-dimensional space to numerically solve Maxwell's equations. This is a direct space time-domain method used for solving electromagnetic problems such as pulsed beam illumination and spectroscopic studies. It solves both electric and magnetic fields using the coupled Maxwell's curl equations by iteration over time. When considering a two dimensional case for the transverse-electric mode TE$_z$ Mode, for example, involving only $E_x$, $E_y$ and $H_z$, the set of time-dependent Maxwell's equations will be [108, 106]:

$$\frac{\partial E_x}{\partial t} = \frac{1}{\varepsilon}\left[\frac{\partial H_z}{\partial y} - \left(J_{\text{source }x} + \sigma E_x\right)\right] \tag{7}$$

$$\frac{\partial E_y}{\partial t} = \frac{1}{\varepsilon}\left[-\frac{\partial H_z}{\partial x} - \left(J_{\text{source }y} + \sigma E_y\right)\right] \tag{8}$$

$$\frac{\partial H_z}{\partial t} = \frac{1}{\mu}\left[\frac{\partial E_x}{\partial y} - \frac{\partial E_y}{\partial x} - \left(M_{\text{source }z} + \sigma^* H_z\right)\right] \tag{9}$$

where $E_x$, $E_y$ is the electric field in $x$ and $y$ directions respectively, $H_z$ is the magnetic field in $z$ direction, $\varepsilon$ is the electrical permittivity, $J_{\text{source}}$ is the electrical current density of the source, $M_{\text{source}}$ is the equivalent magnetic current density of the source, $\sigma$ is the electric conductivity and $\sigma^*$ is the equivalent magnetic loss. This method is particularly useful for direct calculation of nonlinear response of an electromagnetic system. [108]. However, due to its uniform Cartesian grid form for spatial discretization, it limits geometry representation, which results in large memory usage and long computation time. The Finite Element Method (FEM) algorithm is one of the conventional numerical modeling tools for the simulation of near-field problems, as a frequency domain method. It utilizes a mesh made up of tetrahedra, thus combining the advantages of time domain and versatility of spatial discretization procedures and allowing the accurate modeling of complex structures with arbitrarily shaped regions. Furthermore, experimental values of arbitrary dielectric constants for plasmonic micro/nano-structures can be taken into account [109, 108]. To solve the wave equation in terms of the electric field - the Maxwell's curl equation - the FEM method utilizes a vector testing function $v$, integrating on the considered volume [109] For example, for a 2D problem $E_x$, $E_y$ and $H_z$ [104]:

$$\int_\Omega \left[\nabla\left(\frac{1}{\varepsilon_r}\nabla H_z\right) + \frac{\omega^2}{c^2}H_z\right]v\,d\Omega = 0 \tag{10}$$

where $v$ is the test function defined on a domain $\Omega$. For both methods, FDTD and FEM, the perfectly matching layer (PML) method on the outer surface of the computational domain is utilized to absorb waves incident on the boundary [109, 106]. The FIT method covers various electromagnetic problems and can be referred to as a hybrid combination of FDTD and FEM, but differs from these methods by discretization of

Maxwell's equation in an integral form instead of the differential one. For Cartesian grids, the FIT formulation can be rewritten as a standard FDTD method, and for a triangular grid case, the FIT can be linked with FEM method [61, 108, 104].

# 6 What kinds of single cell applications are currently overlooked?

Here we list two applications in which the described super-resolution and optomechanical approaches are essential.

### 6.1 Inflammation

The protein cytokine interleukin IL-1$\beta$ is an inflammation key mediator that is crucial in host-defense responses to infection and injury. IL-1$\beta$ exacerbates damage during chronic disease and acute tissue injury. However, the lack of engineered user-friendly tools for visualizing, monitoring, and manipulating the protein pathways limits investigation of biological processes at the single-cell level in space and time.

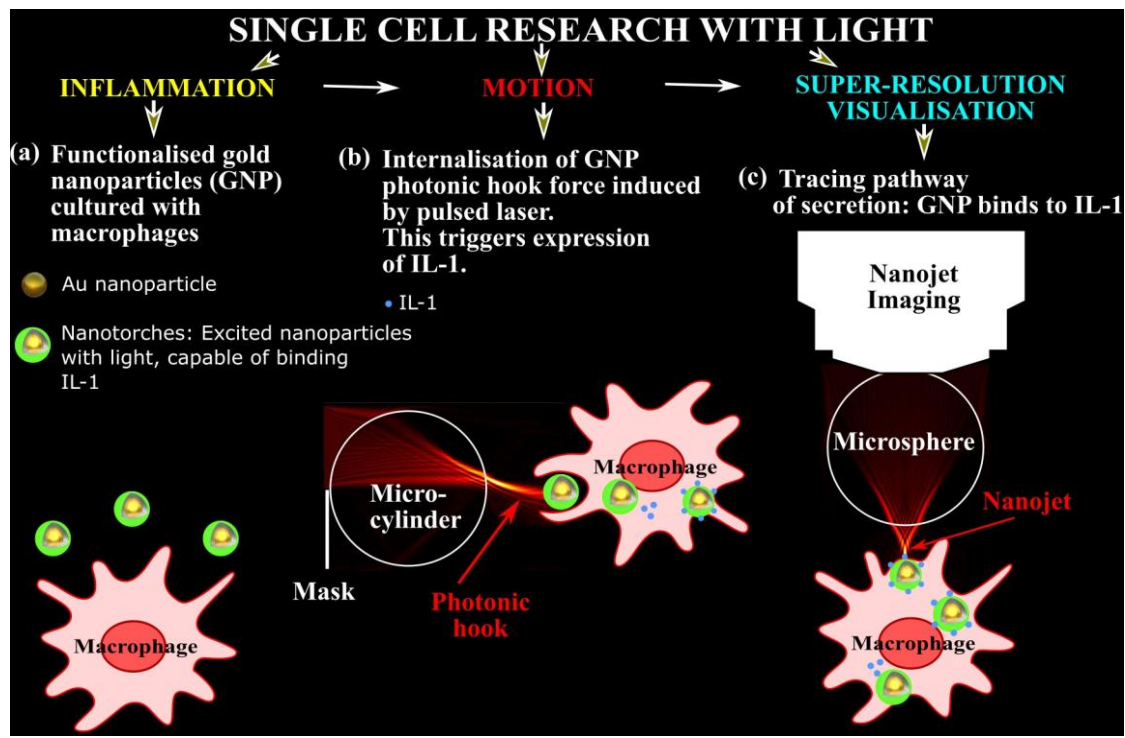

Figure 8: A schematic representation of single-cell process investigation with light (a) biological emulator a cell (b) light- activated tool to initiate internalisation and to control the pathways; (c) tracking, monitoring and nanoscale imaging of proteins or pathways. Here,

functionalised nanotorch is internalised into the intercellular space with the photonic-hook assisted internalisation to trigger the initiation of inflammation cytokine IL-1β, showing calculated repelling curved op- tical field distribution generated with microcylinder, generated IL-1 molecules, and IL-1 connected to nanoparticle. The tracking and nanoscale imaging is performed with the microsphere.

Inflammation alarm cytokine IL-1β is considered an "alarm" upstream mediator which initiates and propagates inflammation. IL-1β is secreted mainly by myeloid cells into the environment immediately after the processing of its precursor, and accumulates in the affected tissues, where it activates diverse cells which abundantly express the IL-1 signaling receptor type 1 (IL-1R1), and collectively contributes to the inflammatory response. The "alarm" function of IL-1β stems from its ability to induce the expression/secretion of pro-inflammatory downstream cascades of cytokines/chemokines, and to also promote cell infiltration through induction of adhesion molecules on endothelial cells and leukocytes. As of now, the majority of IL-1β inhibiting agents - which have also been successfully used clinically in some diseases - neutralize IL-1β already secreted in tissues. These mainly include the IL-1 receptor antagonist (IL-1Ra), which is the physiological inhibitor of IL-1R1 signaling as well as of specific antibodies [110, 2]. However, treatment of patients with such agents is usually applied when IL-1β is already present in the arena and the inflammatory response is already full-blown. IL-1β is active in minute picogram amounts and binds to IL-1R1 at high affinity. Thus, the binding of IL-1β to about as little as 1% of its surface receptors is sufficient to induce a full cellular response. The development of novel therapeutic drugs based on the inhibition of IL-1β secretion would be possible thanks to the new light-based technology.

Inflammation alarm cytokine IL-1β is secreted by the 'leaderless non-vesicular unconventional protein secretion' process. As the name suggests, this process differs from well-understood conventional protein secretion, and the fine details of this process are yet unclear. Advanced light-based technologies may shed light on this process. Unconventional protein secretion is thought to play a role in numerous diseases [111]. For example, the secretion of the HIV-TAT protein by HIV infected cells takes place via an unconventional process and is a crucial step in the pathogenesis of AIDS [112]; unconventional secretion of heat shock proteins plays a crucial role in the immunomodulation and proliferation of cancer [113]; and the unconventional secretion of the FGF2 protein has been postulated as a potential therapeutic target for neurodegenerative diseases such as Alzheimer's and Parkinson's disease and multiple sclerosis [114]. Therefore, the development of tools that could start to unravel the mechanisms of unconventional protein secretion could have a great impact in many areas of disease research and could substantially improve our understanding of this intra-cellular process. IL-1 is an "alarm" cytokine that mediates inflammation, mainly through the induction of a local network of cytokines/mediators. These can serve as potential biomarkers and facilitate cell infiltration into the affected sites through induction of adhesion molecules on endothelial cells and leukocytes [115]. IL-1$\beta$ and IL-1$\alpha$ are the major agonistic molecules of IL-1 and their proteins are encoded by distinct genes that share only slight sequence homology (20-30%) but considerable three-dimensional similarity, which enables them to bind and signal through a common IL-1 receptor type 1 (IL-1R1). IL-1$\alpha$ and IL-1$\beta$ are synthesized as precursors of 31 kD that are further processed by distinct proteases to their mature secreted 17 kD forms. IL-1$\alpha$ and IL-1$\beta$ differ from most other cytokines by lack of a signal sequence, thus not passing through the endoplasmic reticulum-Golgi pathway. This raises intriguing mechanistic, functional and evolutionary questions. In their secreted form, IL-1$\alpha$ and IL-1$\beta$ induce the same biological functions. However, IL-1$\alpha$ and IL-1$\beta$ differ in their compartmentalization within the producing cell or the microenvironment. IL-1$\beta$ is only active in its secreted form and mediates inflammation, which promotes carcinogenesis, tumor invasiveness and immunosuppression [2]. The mechanism of IL-1$\beta$ secretion that is active only upon processing and immediate secretion is not completely understood. It has been suggested that the efflux of calcium into the cell activates phosphatidylcholine-specific phospholipase C and calcium-dependent phospholipase A2, which facilitates the secretion of IL-1$\beta$ together with exocytosis of lysosomal content. The secretion of IL-1$\beta$ through micro-vesicle shedding or via an unknown function of the inflammasome has also been suggested [116]. Recent breakthroughs in inflammasome and IL-1 biology have spurred the development of novel anti-IL-1 agents that are being used in clinical trials in patients suffering from diverse diseases with inflammatory manifestations. IL-1Ra is already FDA-approved and is safe and efficient in alleviating symptoms of rheumatoid arthritis and other diseases characterized by chronic inflammation. In experimental cancer, IL-Ra attenuates tumor-mediated inflammation and invasiveness. IL-1 is abundant in the tumor microenvironment, where it is secreted by the malignant cells, stromal or inflammatory. In experimental tumor models and cancer patients, enhanced expression of IL-1, especially IL-1$\beta$ secreted during tumor progression, has been correlated with a bad prognosis [111]. Understanding the largely unknown mechanisms of IL-1$\beta$ secretion will contribute to better understanding the biology of unconventional protein secretion and to finding new approaches for intervention in the inflammatory process in the body in response to infections, injuries or toxins.

### 6.1.1 The relation between IL-1$\beta$ maturation and its relocation from the cytosol to the plasma membrane

The response to pathogens is orchestrated by the complex interactions and activities of the large num- ber of diverse cell types involved in the immune response. The innate immune response is the first line of defense and occurs soon after pathogen exposure served by phagocytic cells such as macrophages, which carry out this response. Inflammatory macrophages (cells of myeloid lineage) produce an inactive pre- cursor of IL-1$\beta$, named pro-IL-1$\beta$. Following activation of the cytosolic signaling complex named inflammasome [117], activated caspase-1 processes the pro-IL-1$\beta$ to a mature form m-IL-1$\beta$ in the cytosol, which is then relocated to the plasma membrane and secreted [118]. However, the relationship between cytokine processing and secretion is unresolved. We envision the development of tractable light-activated tools to study cellular systems such as inflammatory macrophages - specifically, to adjust light-activated tools to allow an understanding of the relation between IL-1$\beta$ maturation and its relocation from the cy- tosol to the plasma membrane. The experimental strategy may include the creation of a universal plat- form for super-resolution imaging and tracking, and test this platform in studying the relation between IL-1$\beta$ maturation and its relocation from the cytosol to the plasma membrane in pro-inflammatory conditions [119] via expanding light-activated tools [79, 61]. The alternative conventional scheme is a well- studied osmoprotectant glycine to delay caspase-1-dependent macrophage cell rupture, and the resultant passive protein release and monitored IL-1$\beta$ maturation and release upon inflammasome activation in lipopolysaccharide (LPS)-primed macrophages.

Cytokine interleukins (IL-1$\alpha$ and IL-1$\beta$) are the major key mediators in inflammatory response and resistance to pathogens of the innate host immune system. Among them, IL-1$\beta$ is the most powerful cytokine, able to induce inflammatory responses in virtually all tissues of the body [110, 2]. The most dangerous effect of IL-1$\beta$ secretion is the exacerbation of the damage during any chronic disease, and acute tissue injury in cancer and COVID-19, to list a few. Thus, the release mechanisms of IL-1$\beta$ represent a therapeutic target. Consequently, cytokine secretion, in particular IL-1$\beta$, has a huge impact on host wellbeing. Therefore, understanding the secretory pathways of IL-1$\beta$ is critical in understanding inflammation, inflammation-related diseases and resistance to pathogens. However, the mechanism of IL-1$\beta$ release has proven to be elusive. It does not follow the conventional ER-Golgi route of secretion. The important questions of how this protein is exported from cells and how its control affects the inflammatory response remain unanswered. This is partially due to the lack of tools that could allow, in a high- throughput manner, the monitoring, visualization and manipulation of the interleukin pathways. Under- standing long-lasting modifications requires tracking and control of the dynamics of regulatory processes, revealing the mechanism of the inflammatory response and how it resists pathogens. For example, this would require the determination of how a specific secretory pathway leads to enhanced cellular and tissue damage in chronic or acute diseases; how to control it; and when the inflammation begins.

To ensure real-time tracking (monitoring) one may first develop special light sources operating in visible and near-infrared wavelengths and then use these light sources to illuminate a nanoparticle torch (nanotorch) and collect its response; and next, to internalize the nanotorch into a biological emulator macrophage and inspect it with super-resolution visualization tools.

**6.2 Photonic Hook-Controlled Nano-Drug Delivery Across CNS barriers**

A unique characteristic of the central nervous system (CNS) is the existence of physical barriers that keep the microenvironment of the brain and spinal cord in a narrow homeostatic range. Among three barriers - blood-brain (BBB), blood-CSF and arachnoid - the BBB is composed of endothelial cells that are connected by tight junctions and regulate the transport of molecules between the circulatory system and the brain [120]. Yet, while intactness of the BBB is essential for normal brain function, it stands as an obstacle to the permeation of many medicines targeted for the treatment of brain diseases such as tumors [121]. The retina in the eye is an integral part of the CNS, and its blood supply is similar to that of the brain and spinal cord. Indeed, the proper function of the retina requires strict regulation over the molecular composition of the extracellular environment to ensure normal function of the neuronal elements. As an extension of the BBB, the endothelium of the retinal microvessels is interconnected by tight junctions and forms the blood-retinal barrier (BRB) [122]. However, while the brain is protected by the skull and direct access requires surgical intervention, the retina of the eye allows direct visualization. The fact that they share similar properties has increased, in recent years, the idea of the 'eye - a window to brain', i.e. imaging and detection of ocular (retina) abnormalities for diagnosis and monitoring of neurodegenerative diseases [123, 124, 125, 126, 127]. Inversely, several ocular pathologies such as glaucoma [128], age-related macular degeneration [129] and diabetic retinopathy [130] present characteristics of neurodegenerative disorders, including microvascular injury, neuroinflammation and atrophy, that are associated with visual deficits and blindness [131]. Diabetic retinopathy (DR), for example, is a specific complication of diabetes mellitus and is the leading cause of blindness in adults. In DR, hypoxia-associated vascular injury leads to excess production of the vascular endothelial growth factor (VEGF), which further induces abnormal neovascularization and increases the risk to BRB dysfunction and bleeding. Therefore, in recent years, the first-line therapy to DR is the administration of anti-VEGF antibodies. However, like BBB, BRB limits the penetration of antibodies into the retina and forces direct intra-ocular injection, a process that is invasive, expensive and associated with significant complications [132]. The nanojet (photonic-hook), as described above, generates the optomechanical force that allows the motion of nanoparticles. Thus, focal laser illumination of retinal microvessels following non-invasive systemic administration (e.g. intravenous) of VEGF antibody-conjugated nanoparticles has the potential to push the nanoparticle complexes across the BRB, i.e. from the blood into the retina. If this drug delivery approach works, it will allow a non-invasive and efficient treatment for DR.

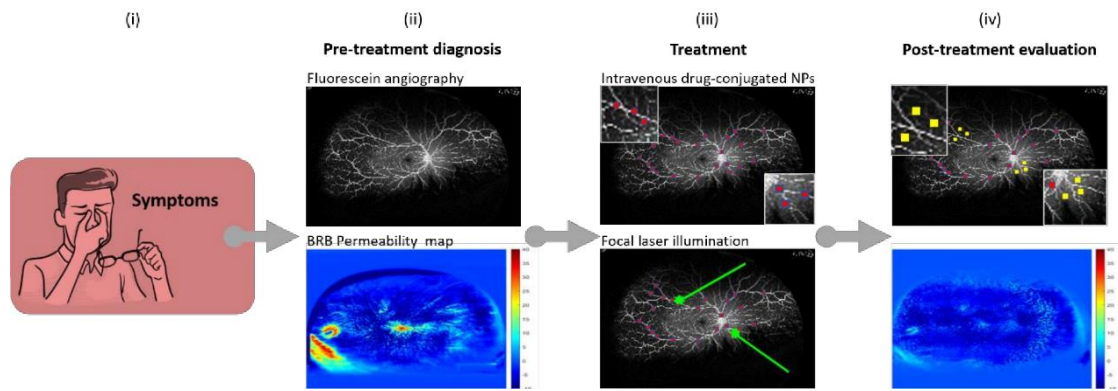

Figure 9: (i) A patient with early stage visual problems and suspected RD will be (ii) referred for fluorescein angiography examination, followed by algorithm-based image analysis for detection of BRB leakiness (pre-treatment). (iii) The patient will be injected intravenously with anti VEGF-conjugated nanoparticles (red spots), followed by a focal laser illumination over areas diagnosed with intact BRB to enhance drug delivery (treatment). (iv) Illumination-induced drug-nanoparticles extravasation (yellow spots - photonic-hook) will result in effective retinal therapy.

## Summary and Future remarks

Despite efforts and advances in engineering light-activated tools for optical visualization at the single-cell level that we overview in this article, a general and robust methodology for investigating and controlling cellular functions by the use of light remains elusive due to fundamental deficits in current methods. These deficits arise from several flaws stemming from current research tools, specifically the nature of 1) optical microscopy; 2) fluorescence microscopy; 3) super-resolution radial fluctuations; 4) optical tweezers. By merging forces into a multidisciplinary research approach for developing light-activated tools, it may become possible to mitigate biological processes such as the inflammatory reaction; to prevent cellular death necrosis in a wide range of diseases such as COVID-19-caused pneumonia, cancer and autoimmune diseases; and to establish new experiments for cellular processes still to be discovered.

Ultimately, light-based techniques may enable the development of a toolbox for the study of secretory pathways of any protein or biomaterial by the use of light, and to generate a novel paradigm for expand- ing the complexity of function that can be imparted to biological systems by leveraging and integrating research in electro-optics, engineering, nanotechnology and biology.

**Supporting Information**

Supporting Information is available from the Wiley Online Library or from the author.

**Acknowledgements**

A.K.* acknowledges the support of the Israel Science Foundation (ISF No. 2598/20)

# References

[1] A. Katiyi, J. Zorea, A. Halstuch, M. Elkabets, A. Karabchevsky, Biosensors and Bioelectronics 2020, 161 112240.

[2] M. Monteleone, A. C. Stanley, K. W. Chen, D. L. Brown, J. S. Bezbradica, J. B. von Pein, C. L. Holley, D. Boucher, M. R. Shakespear, R. Kapetanovic, et al., Cell reports 2018, 24 , 6 1425.

[3] E. Hecht, Optics, 5e, Pearson Education India, 2002.

[4] I. Abdulhalim, A. Karabchevsky, C. Patzig, B. Rauschenbach, B. Fuhrmann, E. Eltzov, R. Marks, J. Xu, F. Zhang, A. Lakhtakia, Applied Physics Letters 2009, 94, 6 063106.

[5] A. Karabchevsky, I. Abdulhalim II, C. Khare, B. Rauschenbach, Journal of Nanophotonics 2012, 6, 1 061508.

[6] Z. Bao, S. Wang, W. Shi, S. Dong, H. Ma, Journal of proteome research 2007, 6, 9 3835.

[7] W. E. Moerner, M. Orrit, Science 1999, 283, 5408 1670.15

[8] X. Li, H. Ma, L. Nie, M. Sun, S. Xiong, Analytica chimica acta 2004, 515, 2 255.

[9] M. Nirmal, B. O. Dabbousi, M. G. Bawendi, J. Macklin, J. Trautman, T. Harris, L. E. Brus, Na-ture 1996, 383, 6603 802.

[10] S. Culley, K. L. Tosheva, P. M. Pereira, R. Henriques, The international journal of biochemistry & cell biology 2018, 101 74.

[11] C. Eggeling, K. I. Willig, S. J. Sahl, S. W. Hell, Quarterly reviews of biophysics 2015, 48, 2 178.

[12] L. Schermelleh, A. Ferrand, T. Huser, C. Eggeling, M. Sauer, O. Biehlmaier, G. P. Drummen, Nature cell biology 2019, 21, 1 72.

[13] E. Wegel, A. Göhler, B. C. Lagerholm, A. Wainman, S. Uphoff, R. Kaufmann, I. M. Dobbie, Scientific reports 2016, 6, 1 1.

[14] E. Betzig, G. H. Patterson, R. Sougrat, O. W. Lindwasser, S. Olenych, J. S. Bonifacino, M. W.Davidson, J. Lippincott-Schwartz, H. F. Hess, Science 2006, 313, 5793 1642.

[15] S. T. Hess, T. P. Girirajan, M. D. Mason, Biophysical journal 2006, 91, 11 4258.

[16] J. S. Biteen, M. A. Thompson, N. K. Tselentis, G. R. Bowman, L. Shapiro, W. Moerner, Naturemethods 2008, 5, 11 947.

[17] M. J. Rust, M. Bates, X. Zhuang, Nature methods 2006, 3, 10 793.

[18] J. P. Pezacki, J. A. Blake, D. C. Danielson, D. C. Kennedy, R. K. Lyn, R. Singaravelu, Naturechemical biology 2011, 7, 3 137.

[19] Y. Park, C. Depeursinge, G. Popescu, Nature photonics 2018, 12, 10 578.

[20] G. Popescu, Quantitative phase imaging of cells and tissues, McGraw-Hill Education, 2011.

[21] A. Ashkin, J. M. Dziedzic, T. Yamane, Nature 1987, 330, 6150 769.

[22] A. Ashkin, J. M. Dziedzic, Science 1987, 235, 4795 1517.

[23] A. Grigorenko, N. Roberts, M. Dickinson, Y. Zhang, Nature Photonics 2008, 2, 6 365.

[24] A. Heifetz, S.-C. Kong, A. V. Sahakian, A. Taflove, V. Backman, Journal of computational and theoretical nanoscience 2009, 6, 9 1979.

[25] L. Yue, B. Yan, Z. Wang, Optics letters 2016, 41, 7 1336.

[26] A. Kovrov, A. Novitsky, A. Karabchevsky, A. S. Shalin, Annalen der Physik 2018, 530, 9 1800129.

[27] J. Y. Lee, B. H. Hong, W. Y. Kim, S. K. Min, Y. Kim, M. V. Jouravlev, R. Bose, K. S. Kim, I.-C. Hwang, L. J. Kaufman, et al., Nature 2009, 460, 7254 498.

[28] Z. Wang, W. Guo, L. Li, B. Luk'yanchuk, A. Khan, Z. Liu, Z. Chen, M. Hong, Nature communications 2011, 2, 1 1.

[29] H. Yang, R. Trouillon, G. Huszka, M. A. Gijs, Nano Letters 2016, 16, 8 4862.

[30] Y. Li, X. Liu, B. Li, Light: Science & Applications 2019, 8, 1 1.

[31] A. Darafsheh, G. F. Walsh, L. Dal Negro, V. N. Astratov, Applied Physics Letters 2012, 101, 14141128.

[32] Y. Geinz, A. Zemlyanov, E. Panina, Russian Physics Journal 2015, 57, 9.

[33] A. Ashkin, Physical Review Letters 1978, 40, 12 729.16

[34] D. Gao, W. Ding, M. Nieto-Vesperinas, X. Ding, M. Rahman, T. Zhang, C. Lim, C.-W. Qiu, Light:Science & Applications 2017, 6, 9 e17039.

[35] A. A. R. Neves, C. L. Cesar, arXiv preprint arXiv:1902.048072019.

[36] C. Bradac, Advanced Optical Materials 2018, 6, 12 1800005.

[37] A. Ashkin, Proceedings of the National Academy of Sciences 1997, 94, 10 4853.

[38] R.-J. Essiambre, Proceedings of the National Academy of Sciences 2021, 118, 7.

[39] I. Minin, Y. E. Geints, A. Zemlyanov, O. Minin, Optics Express 2020, 28, 15 22690.

[40] K. C. Neuman, S. M. Block, Review of scientific instruments 2004, 75, 9 2787.

[41] L. Novotny, B. Hecht, Principles of Nano-Optics, Cambridge University Press, United Kingdom, 2006.

[42] J. Chen, J. Ng, Z. Lin, C. T. Chan, Nat Photon 2011, 5, 9 531.

[43] B. T. Draine, The Astrophysical Journal 1988, 333848.

[44] R. Grimm, M. Weidemüller, Y. B. Ovchinnikov, Advances In Atomic, Molecular, and Optical Physics 2000, 4295.

[45] J. E. Baker, R. P. Badman, M. D. Wang, WIREs Nanomed Nanobiotechnol 2017, n/a–n/a.

[46] A. S. Ang, I. V. Minin, O. V. Minin, S. V. Sukhov, A. S. Shalin, A. Karabchevsky, In META 2018,t he 9th International Conference on Metamaterials, Photonic Crystals and Plasmonics. France, 20182.

[47] J. C. Maxwell, A Treatise on Electricity and Magnetism, Oxford: Clarendon Press, 1873.

[48] P. Lebedev, Ann. Phys. 1901, 311, 11 433.

[49] E. F. Nichols, G. F. Hull, Phys. Rev. (Series I) 1901, 13, 5 307.

[50] P. Lebedev, The Astrophysical Journal 1910, 31385.

[51] A. Ashkin, Phys. Rev. Lett. 1970, 24, 4 156.

[52] A. Ashkin, J. M. Dziedzic, J. E. Bjorkholm, S. Chu, Opt. Lett., OL 1986, 11, 5 288.

[53] A. Dogariu, S. Sukhov, J. Sáenz, Nat Photon 2013, 7, 1 24.

[54] S. Sukhov, A. Dogariu, Rep. Prog. Phys. 2017.

[55] I. Minin, O. Minin, Y. E. Geints, E. Panina, A. Karabchevsky, Atmospheric and Oceanic Optics 2020, 33, 5 464.

[56] L. P. Neukirch, E. Von Haartman, J. M. Rosenholm, A. N. Vamivakas, Nature Photonics 2015, 9, 10 653.

[57] C.-Y. Liu, F.-C. Lin, Optics Communications 2016, 380287.

[58] I. author.(Igor V.) Minin, Diffractive Optics and Nanophotonics: Resolution Below the Diffraction Limit, Springer.

[59] S. Lecler, S. Perrin, A. Leong-Hoi, P. Montgomery, Scientific reports 2019, 9, 1 1.

[60] I. V. Minin, O. V. Minin, C.-Y. Liu, H.-D. Wei, Y. E. Geints, A. Karabchevsky, Optics Letters 2020, 45, 17 4899.17

[61] M. Spector, A. S. Ang, O. V. Minin, I. V. Minin, A. Karabchevsky, Nanoscale Advances 2020, 2, 6 2595.

[62] A. S. Ang, S. V. Sukhov, A. Dogariu, A. S. Shalin, InProgress In Electromagnetics Research Symposium. St. Petersburg, Russia, 2017.

[63] S. Lecler, Y. Takakura, P. Meyrueis, Optics letters 2005, 30, 19 2641.

[64] A. Darafsheh, Journal of Physics: Photonics 2021, 3, 2 022001.

[65] S. Wang, T. Ding, Nanoscale 2019, 11, 19 9593.

[66] B. S. Luk'yanchuk, R. Paniagua-Domínguez, I. V. Minin, O. V. Minin, Z. Wang, Optical Materials Express 2017, 7, 6 1820.

[67] A. Heifetz, J. J. Simpson, S.-C. Kong, A. Taflove, V. Backman, Optics Express 2007, 15, 25 17334.

[68] Y.-C. Li, H.-B. Xin, H.-X. Lei, L.-L. Liu, Y.-Z. Li, Y. Zhang, B.-J. Li, Light: Science & Applications 2016, 5, 12 e16176.

[69] Z. Chen, A. Taflove, V. Backman, Opt. Express, OE 2004, 12, 7 1214.

[70] L. Zhao, C. K. Ong, Journal of Applied Physics 2009, 105, 12 123512.

[71] A. S. Ang, I. V. Minin, O. V. Minin, S. V. Sukhov, A. Shalin, A. Karabchevsky, In Proc. of the the 9th Int. conf. on Metamaterials, Photonic crystals and Plasmonics, Marseille, France, June. 2018.

[72] M. Spector, A. S. Ang, O. V. Minin, I. V. Minin, A. Karabchevsky, Nanoscale Advances 2020, 2, 11 5312.

[73] X. Cui, D. Erni, C. Hafner, Opt. Express, OE 2008, 16, 18 13560.

[74] Y. Li, H. Xin, X. Liu, Y. Zhang, H. Lei, B. Li, ACS Nano 2016, 10, 6 5800.

[75] C. B. Schaffer, A. Brodeur, E. Mazur, Measurement Science and Technology 2001, 12, 11 1784.

[76] A. K. De, D. Roy, A. Dutta, D. Goswami, Applied Optics 2009, 48, 31 G33.

[77] X. Cui, D. Erni, C. Hafner, Optics express 2008, 16, 18 13560.

[78] G. Gu, L. Shao, J. Song, J. Qu, K. Zheng, X. Shen, Z. Peng, J. Hu, X. Chen, M. Chen, et al., Optics express 2019, 27, 26 37771.

[79] A. S. Ang, A. Karabchevsky, I. V. Minin, O. V. Minin, S. V. Sukhov, A. S. Shalin, Scientific reports 2018, 8, 1 1.

[80] L. Yue, O. V. Minin, Z. Wang, J. N. Monks, A. S. Shalin, I. V. Minin, Opt. Lett., OL 2018, 43, 4771.

[81] I. V. Minin, O. V. Minin, Diffractive Optics and Nanophotonics, Springer Briefs in Physics. Springer International Publishing, Cham, 2016.

[82] I. V. Minin, O. V. Minin, G. M. Katyba, N. V. Chernomyrdin, V. N. Kurlov, K. I. Zaytsev, L. Yue, Z. Wang, D. Christodoulides, Applied Physics Letters 2019, 114, 3 031105.

[83] K. Dholakia, G. D. Bruce, Nature Photonics 2019, 13, 4 229.

[84] L. Yue, O. V. Minin, Z. Wang, J. N. Monks, A. S. Shalin, I. V. Minin, Optics letters 2018, 43, 4771.

[85] Y. Zhang, C. Min, X. Dou, X. Wang, H. P. Urbach, M. G. Somekh, X. Yuan, Light: Science & Applications 2021, 10, 1 1.18

[86] L. Kaufman, T. Cooper, G. Wallace, D. Hawke, D. Betts, D. Hess, F. Lagugné-Labarthet, In Plasmonics in Biology and Medicine XVI, volume 10894. International Society for Optics and Photonics, 2019 108940B.

[87] A. Karabchevsky, A. Mosayyebi, A. V. Kavokin, Light: Science & Applications 2016, 5, 11 e16164.

[88] D. R. Dadadzhanov, I. A. Gladskikh, M. A. Baranov, T. A. Vartanyan, A. Karabchevsky, Sensors and Actuators B: Chemical 2021, 333 129453.

[89] D. P. Cherney, J. C. Conboy, J. M. Harris, Analytical chemistry 2003, 75, 23 6621.

[90] I. Prada, L. Amin, R. Furlan, G. Legname, C. Verderio, D. Cojoc, Biotechniques 2016, 60, 1 35.

[91] M. S. Friddin, G. Bolognesi, A. Salehi-Reyhani, O. Ces, Y. Elani, Communications Chemistry 2019, 2, 1 1.

[92] J. Burkhartsmeyer, Y. Wang, K. S. Wong, R. Gordon, Applied Sciences 2020, 10, 1 394.

[93] H. Wang, X. Wu, D. Shen, Optics letters 2016, 41, 7 1652.

[94] M. P. MacDonald, G. C. Spalding, K. Dholakia, Nature 2003, 426, 6965 421.

[95] A. Keloth, O. Anderson, D. Risbridger, L. Paterson, Micromachines 2018, 9, 9 434.

[96] M. Li, T. Lohmuller, J. Feldmann, Nano letters 2015, 15, 1 770.

[97] F. Tian, J. Conde, C. Bao, Y. Chen, J. Curtin, D. Cui, Biomaterials 2016,10687.

[98] R. Weissleder, Nature biotechnology 2001, 19, 4 316.

[99] A. Kusumi, Y. Sako, Current opinion in cell biology 1996, 8, 4 566.

[100] H. Yuan, S. Khatua, P. Zijlstra, M. Yorulmaz, M. Orrit, Angewandte Chemie 2013, 125, 4 1255.

[101] D. R Dadadzhanov, T. A Vartanyan, A. Karabchevsky, Nanomaterials 2020, 10, 7 1265.

[102] A. Karabchevsky, A. Katiyi, A. S. Ang, A. Hazan, Nanophotonics 2020, 9, 12 3733.

[103] D. R. Dadadzhanov, T. A. Vartanyan, A. Karabchevsky, Optics express 2019, 27, 21 29471.

[104] Z. Wang, N. Joseph, L. Li, B. Luk'Yanchuk, Proceedings of the Institution of Mechanical Engineers, Part C: Journal of Mechanical Engineering Science 2010, 224, 5 1113.

[105] M. S. Dhoni, W. Ji, The Journal of Physical Chemistry C 2011, 115, 42 20359.

[106] I. I. Kon, A. Y. Zyubin, A. Seteikin, I. Samusev, In Nanophotonics VIII, volume 11345. International Society for Optics and Photonics, 2020 113452L.

[107] T. Weiland, International Journal of Numerical Modelling: Electronic Networks, Devices and Fields 1996, 9, 4 295.

[108] A. Taflove, S. C. Hagness, M. Piket-May, The Electrical Engineering Handbook 2005, 3.

[109] W. P. Carpes Jr, L. Pichon, A. Razek, International Journal of Numerical Modelling: Electronic Networks, Devices and Fields 2000, 13, 6 527.

[110] R. Sitia, A. Rubartelli, Journal of Biological Chemistry 2020, 295, 22 7799.

[111] J. Kim, H. Y. Gee, M. G. Lee, Journal of cell science 2018, 131, 12.

[112] S. Debaisieux, F. Rayne, H. Yezid, B. Beaumelle, Traffic 2012, 13, 3 355.

[113] T. G. Santos, V. R. Martins, G. N. M. Hajj, International journal of molecular sciences 2017, 18, 5 946.19

[114] M. E. Woodbury, T. Ikezu, Journal of Neuroimmune Pharmacology 2014, 9, 2 92.

[115] A. Rubartelli, F. Cozzolino, M. Talio, R. Sitia, The EMBO journal 1990, 9, 5 1503.

[116] G. Lopez-Castejon, D. Brough, Cytokine & growth factor reviews 2011, 22, 4 189.

[117] S. L. Cassel, F. S. Sutterwala, European journal of immunology 2010, 40, 3 607.

[118] K. Schroder, J. Tschopp, cell 2010, 140, 6 821.

[119] M. S. Boyles, T. Kristl, A. Andosch, M. Zimmermann, N. Tran, E. Casals, M. Himly, V. Puntes, C. G. Huber, U. Lütz-Meindl, et al., Journal of nanobiotechnology 2015, 13, 1 1.

[120] N. J. Abbott, L. Rönnbäck, E. Hansson, Nature reviews neuroscience 2006, 7, 1 41.

[121] C. D. Arvanitis, G. B. Ferraro, R. K. Jain, Nature Reviews Cancer 2020, 20, 1 26.

[122] A. Naylor, A. Hopkins, N. Hudson, M. Campbell, International journal of molecular sciences 2020, 21, 1 211.

[123] N. Cheung, T. Mosley, A. Islam, R. Kawasaki, A. R. Sharrett, R. Klein, L. H. Coker, D. S. Knop-man, D. K. Shibata, D. Catellier, et al., Brain 2010, 133, 7 1987.

[124] M. Koronyo-Hamaoui, Y. Koronyo, A. V. Ljubimov, C. A. Miller, M. K. Ko, K. L. Black, M. Schwartz, D. L. Farkas, Neuroimage 2011, 54 S204.

[125] M. M. Moschos, G. Tagaris, L. Markopoulos, L. Margetis, S. Tsapakis, M. Kanakis, C. Koutsandrea, European journal of ophthalmology 2011, 21, 1 24.

[126] M. L. Monteiro, D. B. Fernandes, S. L. Apóstolos-Pereira, D. Callegaro, Investigative ophthalmology & visual science 2012, 53, 7 3959.

[127] A. London, I. Benhar, M. Schwartz, Nature Reviews Neurology 2013, 9, 1 44.

[128] N. Gupta, Y. H. Yücel, Current opinion in ophthalmology 2007, 18, 2 110.

[129] T. Yoshida, K. Ohno-Matsui, S. Ichinose, T. Sato, N. Iwata, T. C. Saido, T. Hisatomi, M. Mochizuki, I. Morita, et al., The Journal of clinical investigation 2005, 115, 10 2793.

[130] R. Simó, C. Hernández, E. C. for the Early Treatment of Diabetic Retinopathy (EUROCONDOR, et al., British Journal of Ophthalmology 2012, 96, 10 1285.

[131] J. M. Sivak, Investigative ophthalmology & visual science 2013, 54, 1 871.

[132] K. G. Falavarjani, Q. Nguyen, Eye 2013, 27, 7 787.